\documentclass{article}
\usepackage{epsfig}
\usepackage{psfrag}

\usepackage[bottom,hang]{footmisc}
\newenvironment{req}{\setlength\arraycolsep{2pt}\begin{equation}
\begin{array}{rclrclrclrclrclrcl}}{\end{array}\end{equation}}

\newenvironment{reqx}{\setlength\arraycolsep{1.5pt}\begin{eqnarray}}{\end{eqnarray} }

\newcommand{\fig}[1]{Fig.~\ref{#1}}
\newcommand{\eq}[1]{Eq.~(\ref{#1})}
\newcommand{\q}{\qquad}
\newcommand{\qq}{\qquad \qquad}
\newcommand{\f}[2]{\frac{#1}{#2}}
\newcommand{\te}[1]{\textrm{#1}}
\newcommand{\Ri}{\Rightarrow}
\newcommand{\sq}{\sqrt}

\renewcommand{\i}[1]{_{\te{\tiny #1}}}
\renewcommand{\(}{\left(}
\renewcommand{\)}{\right)}
\renewcommand{\.}{\cdot}

\newcommand{\m}\mu{}
\newcommand{\n}{\nu}
\newcommand{\s}{\sigma}
\newcommand{\G}{\Gamma}
\newcommand{\D}{\Delta}

\renewcommand{\l}{\lambda}

\renewcommand{\r}{\rho}     
\renewcommand{\t}{\tau}
\renewcommand{\a}{\alpha}
\renewcommand{\b}{\beta}
\renewcommand{\d}{\delta}

\begin{document}
\noindent
{\Large \bf Embedding Spacetime via a Geodesically Equivalent Metric of Euclidean Signature} 
\\[5mm]
Rickard Jonsson%
\footnote{Department of Astronomy and Astrophysics, Chalmers University of Technology, 
S-412 96 G\"oteborg, Sweden. E-mail: rijo@fy.chalmers.se. Tel +46317723179}
\\
\\
Submitted: 2000-11-06, Published: July 2001\\
Journal Reference: Gen. Rel. Grav. {\bf 33} 1207\\[5mm]
{\bf Abstract.}
Starting from the equations of motion in a 1 + 1 static, diagonal, Lorentzian 
spacetime, such as the Schwarzschild radial line element, I find another metric, but 
with Euclidean signature, which produces the same geodesics $x(t)$. This geodesically 
equivalent, or {\it dual}, metric can be embedded in ordinary Euclidean space. On the 
embedded surface freely falling particles move on the shortest path. Thus one can 
visualize how acceleration in a gravitational field is explained by particles moving 
freely in a curved spacetime. Freedom in the dual metric allows us to display, with 
substantial curvature, even the weak gravity of our Earth. This may provide a nice 
pedagogical tool for elementary lectures on general relativity. I also study extensions 
of the dual metric scheme to higher dimensions. 
\\[5mm]
\\[3mm]


\section{Introduction}
\begin{figure}[b]
\begin{center}
\epsfig{figure=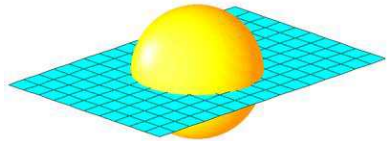,width=5.1cm} 
\caption{A symmetry plane through a star.}
\label{ett}
\end{center}
\end{figure}

\begin{figure}[t]
\begin{center}
\epsfig{figure=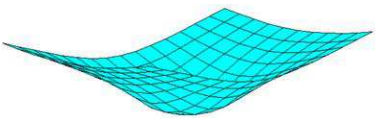,width=5.1cm} 
\caption{The embedding diagram.}
\label{tva}
\end{center}
\end{figure}

It is easy to display the meaning of curved space. For instance we may display the spatial curvature created by a star using an embedding diagram. Figs. \ref{ett}~\&~\ref{tva}.

When it comes to displaying curved spacetime things become more difficult. 
The acceleration of a free test particle is due to a curvature of spacetime. 
What does it mean to have curved time, and can we display it somehow? 

When we create the embedding diagram for the space of a symmetry plane 
through a star, we make a mapping from our spatial plane onto a curved surface 
embedded in Euclidean space. This is done so that distances as measured by 
rulers are the same on the symmetry plane as on the embedding diagram. The 
difference is that on the symmetry plane there is a metric function giving the 
true distance between points whereas in the embedding diagram it is the {\it shape} 
of the surface that gives us the distances. 

If we want to do the same thing for a 1+1 spacetime we immediately run 
into trouble. We have null distances between points and even negative squared 
distances. In the Euclidean space, that we are used to embed in, we can never 
have negative distances.%
\footnote{One {\it can} however embed in a Minkowski space, see Sec \ref{work}.}

\begin{figure}[b]
  \begin{center}
    \psfrag{One to one}{One to one}
    \psfrag{mapping}{mapping}
      	\epsfig{figure=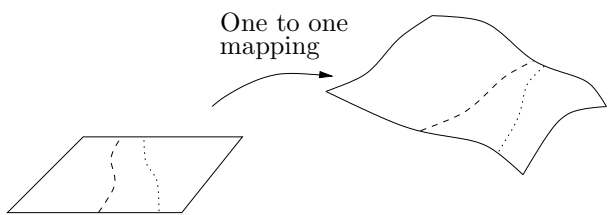,width=8cm}
      	\caption{A mapping to a surface where all freefallers take the shortest path.}  
     	\label{mapping}
  \end{center}  
\end{figure}

Instead of distances, maybe we should study motion. For geodesic lines 
of freefallers in the ordinary spacetime picture of a Schwarzschild black hole 
there is nothing special, or singular, with null geodesics for instance. Would it 
be possible to find a mapping from our coordinate plane to a curved surface such 
that all the worldlines corresponding to test particles in free fall, are geodesics 
(moving on the shortest distance) on the curved surface? See \fig{mapping}.   

Alternatively, if we can not do it for all particles, can we do it for some set 
of geodesics, like photon-geodesics? 

Suppose that we {\it have} found a mapping onto a curved surface such that all 
freefallers move between fixed points like a tightened thread, i.e. on the shortest 
path. Between nearby points on this surface there is then a Euclidean distance as 
measured with a ruler. This means that we can assign a Euclidean distance for 
small displacements in our coordinates. Thus, on our original coordinate plane 
we imagine there to live, not only a Lorentzian%
\footnote{Metrics with both negative and positive distances will be referred to as Lorentzian or simply 
$(+,-)$.}
metric, but also a Riemannian%
\footnote{Metrics with only positive distances will be referred to as Riemannian or simply $(+,+)$. }
metric, that both produce the same geodesics. 

With this understanding, the problem of finding a surface with the right 
curvature is reduced to finding a Riemannian metric that produces the right set 
of geodesics $x(t)$. Once we have found such a geodesically {\it dual} metric, if it 
exists, we can hopefully embed and visualize, a curved spacetime. 

For some special geometries we {\it know} that the dual metric exists. In particular, 
starting from a flat Minkowski spacetime and using standard coordinates, 
the geodesic lines on the coordinate plane are just straight lines. This is also the 
case if we have an ordinary Euclidean metric. In this case we can thus just flip 
the sign of the spatial part of the Minkowski metric to find a dual metric. The 
embedding of a two-dimensional Minkowski spacetime will thus simply be a 
plane.%
\footnote{Or any embedding that is isometric to a plane, for instance a cylinder.}

Also, in a general Lorentzian spacetime, we can at every point choose coordinates 
so that the metric reduces to Minkowski, with vanishing derivatives. 
In this local, freely falling, coordinate system particles move on straight lines. 
This means that they move on the longest path in the local Lorentzian spacetime. 
However they also move on the {\it shortest} path in the corresponding local 
Euclidean spacetime. Then we know that there exists a dual metric that 
produces the right equations of motion at least in every {\it single} point. The question 
is whether we can connect all these single point metrics in a smooth way. 

Notice that constant time lines for inertial observers in Minkowski, are 
straight lines. They are also geodesics, moving the longest path, if we change 
the sign of the whole metric so that spacelike distances becomes positive.%
\footnote{ In this article we use the convention that squared timelike distances are positive. 
Alternative to flipping the sign of the entire metric we can say that the imaginary 
distance traveled is maximized.}
This means that if we find a dual metric, there will be geodesics that correspond to 
local time lines of freefallers. We may also consider these lines to be particles 
moving faster than light, so called tachyons. 

On the curved surface there is in principle no way to distinguish between 
timelike and spacelike displacements from the {\it shape} of the surface. However, 
on the surface lives the original Lorentzian metric that tells us the true distances 
between nearby points. In other words there are small Minkowski systems living 
on the surface, telling us the proper distance between points. 

In a sense all we are doing is {\it shaping} the manifold. We still need the ordinary 
metric to get distances right. The difference is that we do not need this 
function to get the {\it geodesics} right. This follows from the shape of the surface.


\section{The dual metric in $1+1$ dimensions}
Let us for simplicity, start the analysis with a 1+1 time independent 
diagonal metric, with Lorentzian signature, and see if we can find a time independent 
and diagonal dual metric with Euclidean signature.


\subsection{Equations of motion}
Assume that we have a line element:
\begin{eqnarray}
d\t^2=a(x) \. dt^2 +c(x) \. dx^2
\end{eqnarray}
Using the squared Lagrangian formalism (see e.g. \cite{dinverno}), 
we immediately get the integrated equations of motion:
\begin{eqnarray}
a \(\f{dt}{d\t}\)^2 + c \(\f{dx}{d\t}\)^2&=&1\\
a \f{dt}{d\t} &=& K \label{eq3}
\end{eqnarray}
Here $K$ is a constant for every geodesic.%
\footnote{Notice that for spacelike geodesics, 
i.e. tachyons or lines of constant time, we have an imaginary $K$ for the ordinary Schwarzschild metric.} 
Now we want to find an equation for $x(t)$. Introducing $\s=1/K^2$ for compactness, the result is:
\begin{eqnarray}
\(\f{dx}{dt}\)^2=\f{a}{c} \(\s \  a - 1\) \label{xprim}
\end{eqnarray}
Notice that this equation applies to both $(+,-)$ and $(+,+)$ metrics.


\subsection{The dual metric equations}
The question is now: What other metrics, if any, can produce the 
same set of geodesics $x(t)$? Denoting the original metrical components 
by $a_0$ and $c_0$ and the original constant of the motion for a certain 
geodesic by $\s_0$ we must have:
\begin{eqnarray}
\f{a}{c} (\s a - 1)= \f{a_0}{c_0}(\s_0 a_0 - 1) \label{master}
\end{eqnarray}
This relation must be fulfilled for every $x$ and every geodesic.
Notice that $\s$ may depend on $\s_0$ only.

We see immediately that we can regain the old metric simply by 
setting $a=a_0$, $c=c_0$ and $\s(\s_0)=\s_0$. 
As for other solutions they may appear hard to find at first. 
We know however that starting from e.g. Minkowski, 
we must be able to flip the sign on the spatial part of the metric, 
without affecting the geodesic lines. Let us however rewrite \eq{master} a bit:
\begin{eqnarray}
\s= \( \f{c}{c_0}\f{a_0^2}{a^2}\) \.\s_0 + \(\f{1}{a}-\f{c}{c_0}\f{a_0}{a^2}\)
\end{eqnarray}
Now we see more clearly that if this relation is to hold for all $x$ and all $\s_0$ we must have:
\begin{eqnarray}
k_1=\f{c}{c_0}\f{a_0^2}{a^2}  \qq
k_2=\f{1}{a}-\f{c}{c_0}\f{a_0}{a^2}  \label{kkk2}
\end{eqnarray}
We have thus a linear relation between the constants of the motion:
\begin{eqnarray}
\s=k_1 \. \s_0 + k_2  \label{sigma}
\end{eqnarray}
Here $k_1$ and $k_2$ are constants that depend on neither $x$ nor $\s_0$. From \eq{kkk2} we may solve for $a$ and $c$ in terms of $k_1$ and $k_2$:
\begin{eqnarray}
a=\f{a_0}{a_0 k_2 + k_1} \qq
c=\f{c_0 k_1}{\(a_0 k_2 + k_1\)^2}
\end{eqnarray}
Defining $\a=1/k_2$ and $\b=-k_1/k_2$ this may be rewritten as:
\begin{eqnarray}
a&=&\a \f{a_0}{a_0 - \b}\label{a} \\[2mm] 
c&=&\a \f{-c_0 \b}{\(a_0 -\b\)^2} \label{c}
\end{eqnarray}
We see that $\a$ is a pure scaling constant whereas $\b$ is connected to compression along the $x$-axis as will be discussed later. 

Now the question is: can we choose $\a$ and $\b$ so that, assuming a Lorentzian 
original metric (positive $a_0$ and negative $c_0$) we get a Riemannian dual metric? 
Indeed necessary and sufficient conditions for the dual metric to be positive definite is:%
\footnote{It might appear that we would get extra restrictions on these constants from demanding that $\s$ in \eq{sigma} must be positive. This constraint turns out to be identical to the constraint of \eq{conbeta} however.}
\begin{eqnarray}
&&0<\a \\[2mm]
&&0<\b<a_0  \label{conbeta}
\end{eqnarray}


\subsection{The Schwarzschild exterior metric}
Introducing dimensionless and rescaled coordinates $x=\f{r}{2MG}$  and 
correspondingly rescaling Schwarzschild and proper time, the ordinary Schwarzschild 
metric is given by: 
\begin{eqnarray}
a_0=\(1-\f{1}{x} \) \qq
c_0=-\(1-\f{1}{x} \)^{-1}
\end{eqnarray}
At spatial infinity this reduces to $a_0=1$ and  $c_0=-1$. At infinity our new metric is thus reduced to:
\begin{eqnarray}
a_{\infty}=\a \f{1}{1 - \b} \qq
c_{\infty}=\a \f{\b}{\(1 -\b\)^2}
\end{eqnarray}
Let us study the quotient of $a_{\infty}$ and  $c_{\infty}$:
\begin{eqnarray}
\f{a_{\infty}}{c_{\infty}}=\f{1}{\b} -1
\end{eqnarray}
We see that for $\b>1/2$ the quotient is smaller than 1. This implies a stretching in $x$. 
When we embed our new metric this will correspond to opening up the photon lines so they 
become more parallel to the constant time line. 

In particular, demanding that $a_{\infty}=1$ and  $c_{\infty}=1$ yields $\a=1/2$ and $\b=1/2$. 
Using this particular {\it gauge}, from now on denoted the {\it standard} gauge, 
we get from \eq{a} and \eq{c} the dual line element:
\begin{eqnarray}
ds^2=\f{x-1}{x-2} \cdot dt^2+\f{x^3}{(x-1)(x-2)^2} \cdot dx^2
\end{eqnarray}
This metric has positive metrical components from infinity and in to $x=2$. We 
have thus succeeded, and found a Riemannian dual metric, at least on a large 
section of the spacetime.


\subsection{On the interpretation of $\a$ and $\b$} \label{duality}
We may rewrite \eq{a} and \eq{c} as:
\begin{eqnarray}
\f{a}{\a}=\f{a_0/\b}{a_0/\b - 1} \qq 
\f{c}{\a}=\f{-c_0/\b}{\(a_0/\b -1\)^2} 
\end{eqnarray}
Then we see that just as $\a$ is a rescaling of the new metric -- so is $\b$ a rescaling of 
the {\it original} metric. We can easily work out the {\it inverse} of the relations above to find:
\begin{eqnarray}
\f{a_0}{\b}=\f{a/\a}{a/\a - 1} \qq
\f{c_0}{\b}=\f{-c/\a}{\(a/\a -1\)^2} 
\end{eqnarray}
We see that we have a perfect symmetry in going from the original metric to the dual and vice versa, 
justifying the duality notion. This symmetry would be even more obvious if 
we would denote $\b$ by $\a_0$ instead.

With hindsight we realize that we must have two rescaling freedoms just 
like that. A metric that is dual to some original metric must also be dual to a 
twice as big original metric and vice versa. Notice however that if we make the 
original space twice as big, then the dual space does not automatically become 
twice as big. It {\it does} however if we double both $\a$ and $\b$!


\section{The embedding equations}
\begin{figure}[b]
  \begin{center}
    \psfrag{r}{$r$}
    \psfrag{z}{$z$}
      	\epsfig{figure=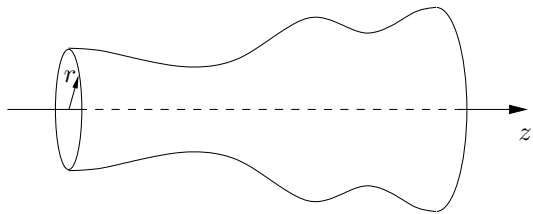,width=7cm}
      	\caption{A rotational surface.}  
     	\label{randz}
  \end{center} 
\end{figure}

In general if we have a metric with a symmetry, we can embed it as a 
rotational surface if we can embed it at all. Our task is then to find a 
radius $r(x)$ and a height $z(x)$ for the embedding of the dual metric. See \fig{randz}.

\subsection{Finding $r(x)$} 
For pure $t$-displacements the dual distance traveled is $ds=\sq{a}\.dt$. We realize that we must have:
\begin{eqnarray}
r=k \sqrt{a}    \label{k}
\end{eqnarray}
Here $k$ is a constant of the embedding only, it does not affect the way we measure 
distances on the surface -- only its {\it shape}. See Section \ref{shape}. 
In terms of the original metric:
\begin{eqnarray}
r=k \. \sq{\a \f{a_0}{a_0 - \b} } \label{r}
\end{eqnarray}
In particular for the Schwarzschild case using the standard gauge ($\a=1/2$, $\b=1/2$), and $k=1$:
\begin{eqnarray}
r(x)=\sq{\f{x-1}{x-2}}  \label{seer}   
\end{eqnarray}
We see that as $x$ tends to infinity we have a unit radius, 
whereas approaching $x=2$ from infinity the radius blows up. Already now we may understand the 
qualitative behavior of the embedding diagram. The curious reader may jump 
immediately to Section \ref{embedit}.


\subsection{Finding $z(x)$}
From \fig{dzdx} using the Pythagorean theorem we find:
\begin{figure}[b]
  \begin{center}
    \psfrag{dl}{$dl=\sqrt{c} \cdot dx$}
    \psfrag{dz}{$dz$}
    \psfrag{dr}{$\displaystyle \frac{dr}{dx} \cdot dx$}
      	\epsfig{figure=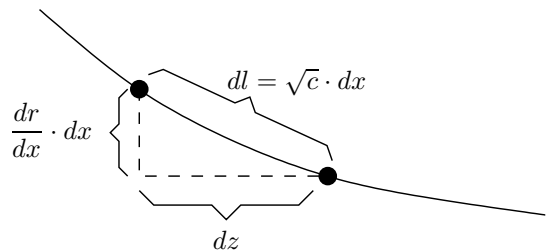,width=7.2cm}
      	\caption{The relation between $dz$, $dr$ and $dl$.}  
     	\label{dzdx}
  \end{center} 
\end{figure}
\begin{eqnarray}
dz=dx \. \sq{c(x)-\(\f{dr(x)}{dx}\)^2}
\end{eqnarray}
Using \eq{c} and \eq{r} and defining ${a'_0}=da_0/dx$ we readily get:
\begin{eqnarray}
dz=dx \. \sq{\a \f{-c_0 \b}{(a_0-\b)^2}  -    \f{k^2 \a \b^2}{4} \. \f{{a'_0}^2}{a_0 (a_0-\b)^3}}
\end{eqnarray}
So:
\begin{eqnarray}
\D z=\sq{\a}\sq{\b} \int dx \sq{\f{-c_0}{(a_0-\b)^2}  -    \f{k^2 \b}{4} \. \f{{a'}_0^2}{a_0 (a_0-\b)^3}} \label{dz}
\end{eqnarray}


\subsection{Embedding criterions} \label{embedding}
We see from \eq{dz} that there is a limit as to how big the embedding constant $k$ can 
be lest we get something negative within the root:
\begin{eqnarray}
k^2<\te{min} \left\{-c_0 \f{4}{\b} \f{a_0 (a_0-\b)}{(a'_0)^2}\right\} \label{embed}
\end{eqnarray}
Let us investigate what this restriction amounts to for the specific cases of the 
Schwarzschild exterior and interior metric.


\subsubsection{The exterior metric}
For the exterior Schwarzschild metric, the expression within the brackets of 
\eq{embed} is smaller the closer to the gravitational source that we are, approaching
0 before we reach the horizon. On the other hand it goes to infinity as $x$ goes to infinity, 
since $a'_0$ goes to zero here. This means that, for any given $k$, the embedding {\it works}, 
as we approach infinity. It however only works from a certain point in $x$ and onwards.

For the exterior Schwarzschild we have:
\begin{eqnarray}
a_0=1-\f{1}{x} \qq a'_0=\f{1}{x^2} \qq -c_0=\(1-\f{1}{x}\)^{-1}
\end{eqnarray}
After some minor juggling we then find:
\begin{eqnarray}
k<\f{2x^2}{\sq{\b}} \sq{1-\b-\f{1}{x}}
\end{eqnarray}
Assuming that we are using the standard gauge, $\b=1/2$, and $k=1$ this reduces to a restriction in $x$:
\begin{eqnarray}
x^3(x-2)>\f{1}{4}  \q \Ri \q  x>2,02988...
\end{eqnarray}
So, using the standard gauge, the dual embedding only exists from roughly two Schwarzschild radii 
and on towards infinity. Incidentally may insert this innermost $x$ into \eq{seer} and find $r\sim 5.87$.

Notice that these numbers only apply to the particular boundary condition 
where we have Pythagoras and unit embedding radius at infinity. By choosing 
other gauge constants and embedding constants we can embed the spacetime as 
close to the horizon as we want. Notice that there is nothing physical with the 
limits of the dual metric and the embedding limit. They are merely unfortunate 
artifacts of the theory.


\subsubsection{The interior metric of a star}
Assuming a static, spherically symmetric star consisting of a perfect fluid 
of constant proper density, we have the standard Schwarzschild interior metric: 
\begin{eqnarray}
a_0&=&\f{1}{4}\(3 \sqrt{1-\f{1}{x_0}}-\sqrt{1-\f{x^2}{x_0^3}}\)^2\\
c_0&=&-\(1-\f{x^2}{x_0^3}\)^{-1}
\end{eqnarray}
Here $x_0$ is the $x$-value at the surface of the star.
For the interior star the embedding criterion, \eq{embed}, becomes:
\begin{eqnarray}
k^2<\f{4 x_0^6}{\b} \te{min} \left\{\f{1}{x^2}\.\(a_0-\b\)\right\} \label{kk}
\end{eqnarray}
The bracketed function can be either monotonically decreasing, have a local minima within the star or even be monotonically increasing depending on $x_0$ and $\b$.

Apart from the embedding restriction we have of course the restriction on 
the metric itself. Since $a_0$ is monotonically increasing, for interior plus exterior 
metric, the metrical restriction for the entire star becomes: 
\begin{eqnarray}
\b<a_0(0)
\end{eqnarray}
Using $\b=1/2$ and the interior $a_0$ in the center of the star this restriction becomes 
a restriction in $x_0$:
\begin{eqnarray}
x_{0}>\f{9}{9-(\sq{2}+1)^2}\simeq 2,838
\end{eqnarray}
For any $x_0>2,837$  and  $\b=1/2$ the right hand side of \eq{kk} 
is always considerably larger than $1$ and thus the embedding imposes no extra 
constraints on $x_0$ in the gauge in question, assuming $k=1$. 
Incidentally we see from \eq{r} that the radius of the central bulge goes to infinity 
as $x_0$ approaches its minimal value.


\subsection{On the interpretation of the embedding constant $k$} \label{shape}
\begin{figure}[b]
  \begin{center}
    \psfrag{Roll tighter}{Roll tighter}
    \psfrag{k=1}{$k=1$}
    \psfrag{k<1}{$k<1$}
    \epsfig{figure=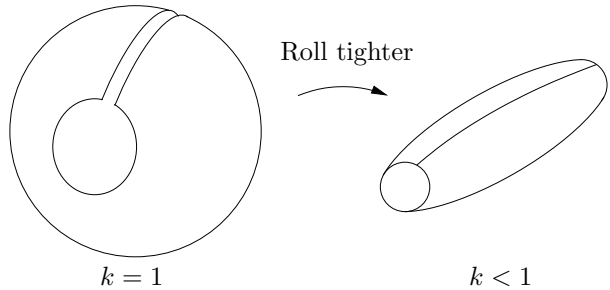,width=8.3cm}
      	\caption{Two isometric embeddings of a sliced-open sphere.}  
     	\label{sphere}
  \end{center} 
\end{figure}

Recall \eq{k}:
\begin{eqnarray}
r(x)=k \sqrt{a(x)}
\end{eqnarray}
We see that $k$ determines the scale for the embedding radii. Notice however 
that the distance to walk between two radial circles, infinitesimally displaced, 
is determined by $dl=\sqrt{c}dx$. So, when we double $k$ we double the radius of 
all the circles that make up the rotational body while keeping the distance to 
walk between the circles unchanged. This means that we increase the slope of 
the surface everywhere. The more slope the bigger the increase of the slope. If 
we increase $k$ too much the embedding will fail. An example of how different 
$k$ affects a certain embedding is given in \fig{sphere}.

Now consider \eq{r}:
\begin{eqnarray}
r(x)=k \sq{\a} \sqrt{\f{a_0}{a_0-\b}}
\end{eqnarray}
We notice that, while 
increasing $k$ we can {\it decrease} $\a$ in such a way that we do not change {\it any} $r(x)$. 

The net effect on the embedding is then to compress the 
surface in the $z$-direction, while keeping all radii. This is done in such a way that 
all distances $(dl)$ on the surface in the $z$-direction is reduced by the same factor 
everywhere. This means that where the slope is big we compress a lot in the 
$z$-direction. Also since we are rescaling the dual metric, meter- and second-lines 
on the surface will move closer. 
Using this scheme we can produce substantial curvature out of something 
that was originally almost cylindrical. Also, using our $\b$-freedom, we can flip 
down the photon lines towards the time line to better suit what we humans 
experience. This way we have a chance of displaying, with reasonable distances and 
curvatures, why things accelerate at the surface of the Earth. See Section \ref{earth}.


\section{The embedding diagram}\label{embedit}
\begin{figure}[b!]
  \begin{center}
    \psfrag{Towards spatial}{Towards spatial}
    \psfrag{infinity}{infinity}
    \psfrag{In}{In}
    \psfrag{Out}{Out}
    \psfrag{Freely falling observer}{Freely falling observer}
    \psfrag{moves out and back}{moves out and back}
    \psfrag{Time}{Time}
    \psfrag{Observer}{Observer}
    \psfrag{at rest}{at rest}
    \epsfig{figure=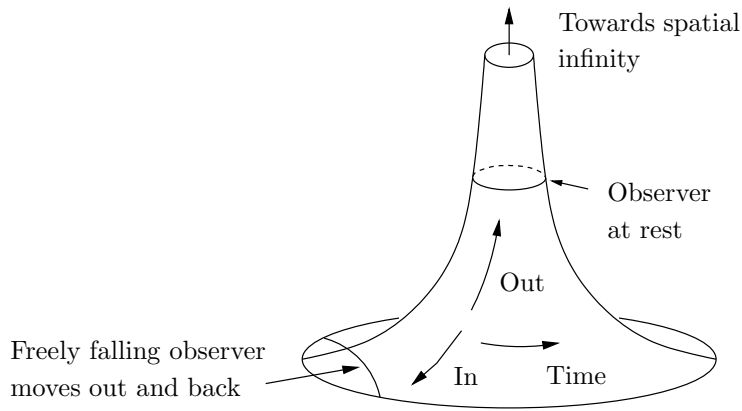,width=10cm}
      	\caption{A vision of spacetime.}  
     	\label{tophat}
  \end{center} 
\end{figure}

Already from  \eq{seer}  we realize quantitatively how the new dual 
$(+,+)$ spacetime must look like. See \fig{tophat}. Notice that time is the azimuthal angle and 
the whole spacetime is layered (infinitely thin), like a toilet roll. 

The geometry will approach a cylinder as we go towards spatial infinity. 
This is as it should be since we want a flat%
\footnote{A cylinder is an intrinsically flat geometry.}
spacetime where there is no gravity. 

Notice that in an ordinary embedding of an equatorial plane of a black hole, 
the geometry opens up towards infinity and the little hole is at the horizon. Here 
it opens up towards the horizon and the little hole is towards infinity. 

So, we have found a dual $(+,+)$ spacetime of 
Schwarzschild, that can be embedded, where {\it all} particles move on geodesics, 
i.e. {\it shortest} distance. We can thus make a real model, say in 
polished metal, and then find possible geodesics just by tightening a 
thin thread between pairs of points on the surface.

Alternative to tightening threads, one could put a little toy car or 
motorcycle, on the surface. Starting the car at some point, directed solely in the azimuthal 
direction, and pushing the car straight forward will result in a spiral inwards. 
Thus we see how moving straight forward can result in acceleration. Also, if we 
want the car to stay at a fixed $x$ we notice that we must turn the wheel (assuming 
an advanced toy car), so that the car is constantly turning left (e.g.), i.e. 
accelerating upwards. This illustrates in an excellent manner how it is possible for us 
Earthlings to always accelerate upwards without ever going anywhere. 

\begin{figure}[b!]
  \begin{center}
    \psfrag{Observer oscillating around}{Observer oscillating around}
    \psfrag{the center of the star}{the center of the star}
    \psfrag{Observer at rest at}{Observer at rest at}
    \psfrag{the center of the star}{the center of the star}
    \psfrag{Time}{Time}
    \psfrag{Constant time line}{Constant time line}
    \psfrag{Space}{Space}
    \psfrag{mi}{$-\infty$}
    \psfrag{pi}{$+\infty$}
        \epsfig{figure=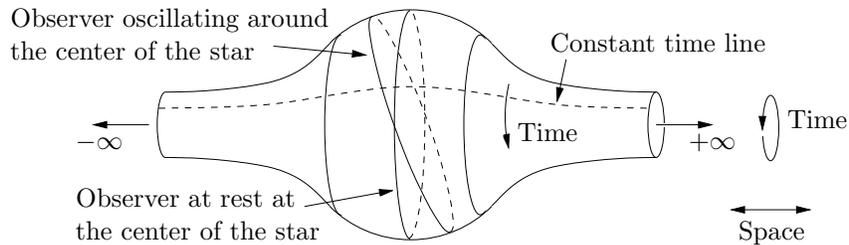,width=10.9cm}
      	\caption{The spacetime of a central line through a star.}  
     	\label{star}
  \end{center} 
\end{figure}

Now that we have understood the name of the game in this embedding 
scheme we can figure out qualitatively how the embedded spacetime of a line 
{\it through} a star must look like. This is depicted in \fig{star}. For better layout, 
and also to more naturally connect the embedding diagram to the ordinary 
Schwarzschild diagram, we have now {\it space} in the left-right direction. 

A particle oscillating around the center of the star is nothing but a 
thread winding around the central bulge. Notice however that we do not generally expect to 
have something close to a sphere for the interior embedding.
For a non-compact star we would rather expect%
\footnote{This is actually not obvious however -- and not even always the case as we will understand later.}
something close to a cylinder, with a long slightly bulged interior star.
Also, if we would have a perfect sphere for the the interior, 
then oscillations around the center of the star would correspond to 
great circles. This would mean that the period of revolution, as measured in 
Schwarzschild time, would be independent of the amplitude of the oscillation. 
This is actually true, for a constant density star, in Newtonian theory. In the full 
theory, and for more general density distributions, we will not expect a perfect 
sphere however. Also, more embeddings than the sphere has the focusing feature 
that makes the period of revolution independent of the amplitude. 

\begin{figure}[t]
  \begin{center}
    \psfrag{Flying saucer shaped geometry}{Flying saucer shaped geometry}
    \psfrag{Geometry seen from the side}{Geometry seen from the side}
    \psfrag{with winding geodesics}{with winding geodesics}
    \psfrag{Small amplitude}{Small amplitude}
    \psfrag{Large amplitude}{Large amplitude}
      	\epsfig{figure=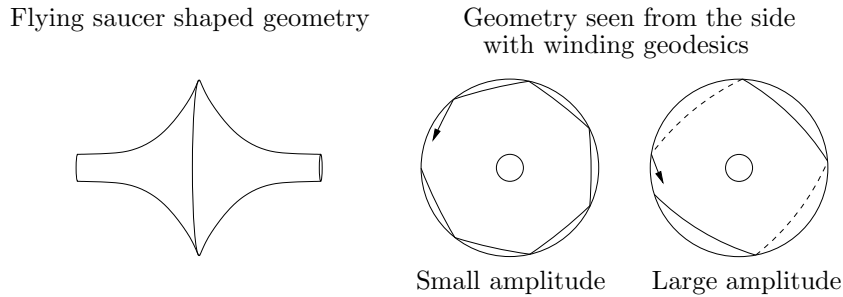,width=10.9cm}
      	\caption{How the shape of the bulge, due to the density distribution, 
       affects the the amplitude dependence of the periods of revolution for 
       freefallers around the center of the star.}  
     	\label{density}
  \end{center} 
\end{figure}

How a certain density can affect the shape of the interior bulge is depicted 
in \fig{density}. 

In this geometry it is obvious that increasing amplitude means increasing 
period of revolution. This is exactly what may be expected from Newtonian 
theory if the density is increasing towards the center. We may also consider 
the opposite situation with decreasing density in the center of the star. Then 
the central parts of the embedding will be close to cylindrical, and it is easy to 
imagine that increasing amplitude means decreasing period of revolution. Notice 
however that if we want to find the exact dependence of $x(t)$, from the embedding 
diagram, we need also to know how the $x$'s are distributed on the surface. 

It is fascinating that we can {\it visualize} how density creates spacetime curvature,
which in turn affects the geodesics of particles


\subsection{Numerically calculated diagrams}
\begin{figure}[tb]
  \begin{center}
      	\epsfig{figure=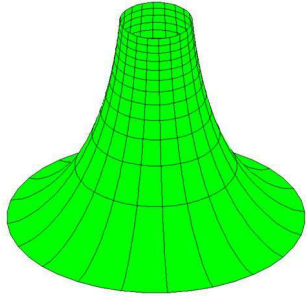,width=4.1cm}
      	\caption{A standard embedding of the exterior spacetime of a star for $2.03<x<3.00$. 
                 The spatial lines are equidistant in $x$ with a spacing that is one fourth 
                 of the spacing between the time lines (for esthetical reasons).}  
     	\label{newext}
  \end{center} 
\end{figure}

Numerically it is no problem to integrate \eq{dz}. For the exterior metric a 
particular result is depicted in \fig{newext}.

\begin{figure}[b!]
  \begin{center}
      	\epsfig{figure=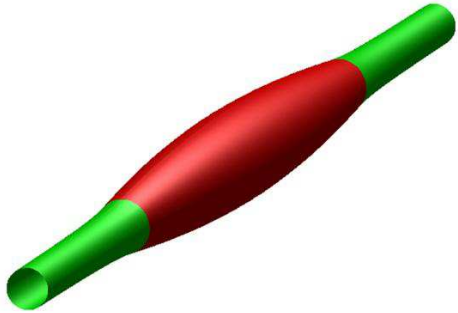,width=6.2cm}
      	\caption{A standard embedding of the spacetime of a star with $x_0 \sim 3.2$.}  
     	\label{num2}
  \end{center} 
\end{figure}

We may as easily get the embedding for the full star. One result of this, 
where we have omitted the coordinate lines, is depicted in \fig{num2}. 

One may reflect that the embedding is not as bulgy as expected, and still I 
have chosen a compactness for which it is about as bulgy as it gets for a standard 
embedding. 

The main reason for the flatness is that as one moves towards the center 
of the star and the embedding radius is increasing, photon geodesics (and other 
geodesics) will be tilted further towards the constant time line. This is a direct 
effect of photons moving the shortest distance on the rotational surface 
(see Section \ref{rotation}). Also as the radius increases, moving in Schwarzschild time, 
means moving a longer distance on the surface, remember that the Schwarzschild time is 
proportional to the azimuthal angle. These two radial effects means that $dz/dt$, 
for photon geodesics, increases with increasing $r$. We therefore understand that 
we {\it must} stretch the bulge in the $z$-direction to insure that photons do not pass 
the star too quickly.

If we want a star with more shape we can increase the $k$-value. This 
increases the tilt of the surface everywhere, making the embedding bulgier while 
photon lines at infinity remains at $45^o$.


\section{The weak field limit}\label{weak}
The dual metric and the embedding formulae are rather mathematically 
complicated, especially for the interior star. To gain some intuition it will prove 
worthwhile to study the weak field limit, where we can Taylor expand our 
expressions. In this section we will not use rescaled coordinates, $x$, but the 
ordinary radius, denoted by $\r$ so as not to confuse it with the embedding radius $r$. 
Let us use the standard gauge $\a=\f{1}{2}$, $\b=\f{1}{2}$ and also $k=1$ for simplicity. We define:
\begin{eqnarray}
a_0=1-\varepsilon (\r)
\end{eqnarray}
Assuming $\varepsilon(\r)$ to be small we may Taylor expand the expression for the embedding radius:
\begin{eqnarray}
r&=&k \sqrt{\a} \sqrt{\f{a_0}{a_0-\b}} \q \Ri \q r\simeq 1+\f{\varepsilon}{2} 
\end{eqnarray}
Introducing $r=1+h$, we have thus to lowest order $h=\varepsilon/2$.   
One may show \cite{weinberg} that for a stationary, weak field in general we have:
\begin{eqnarray}
a_0=(1+2 \. \phi)
\end{eqnarray}
Here $\phi$ is the dimensionless (using $c\i{light}=1$) Newtonian potential per unit mass,
e.g. $GM/\r$ for a point mass. We thus conclude that:
\begin{eqnarray}
h=-\phi  \label{h}
\end{eqnarray}
So in first order theory, using the standard gauge and $k=1$, the height of the 
perturbation equals exactly minus the Newtonian potential.%
\footnote{If we are starting from the mass rescaled metric --
the height of the perturbation at any $x \sim z$ will be the dimensionless 
Newtonian potential per unit mass divided by the mass of the gravitating system.}
This result is actually not to surprising. See section \ref{rotation}.

Incidentally we may also show that:
\begin{equation}
\begin{array}{rclrcl}
\begin{array}{ll}
a_0=1-\varepsilon (\r)  \\[-2mm] 
c_0=-(1+\d (\r)) \end{array} \q \Ri \q z\simeq \int d\r \  1 + \displaystyle \f{\d(\r)}{2} + \displaystyle \f{\varepsilon(\r)}{2}
\end{array}
\end{equation}


\subsection{Applications}
\begin{figure}[h!]
  \begin{center}
      \epsfig{figure=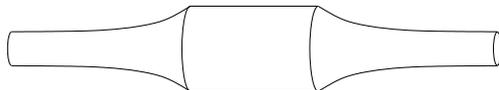,width=6.6cm}
      \caption{The rolling-pin spacetime of a central line through a 
               Newtonian shell of matter.}  
      \label{broedkavle}
  \end{center} 
\end{figure}

We may use our newly found intuition from Newtonian theory to create a new interesting picture. 
Suppose that we have a static spherical shell of some mass. 
Inside the shell we have no forces and thus $\phi$ is constant. 
According to the derivation above we would then have constant $h$ in the interior of the star. 
See \fig{broedkavle}.

We see that inside the shell the geometry is flat, consistent with having no gravitational forces.

\begin{figure}[h!]
  \begin{center}
      \epsfig{figure=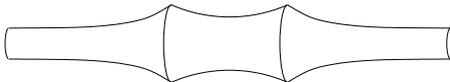,width=6.0cm}
      \caption{The spacetime of a central line through a certain energy distribution.}  
      \label{kassimir}
  \end{center} 
\end{figure}

I will leave to the reader to figure out what strange {\it spherical} mass distribution 
that could give rise the the embedding diagram depicted in \fig{kassimir}.

\section{On geodesics on rotational surfaces} \label{rotation}
Since this paper utilizes geodesics on rotational surfaces -- maybe a general note 
on the subject is in order. Parameterizing any rotational surface with $r$ and $\varphi$, 
the metric can be written (in every region of monotonically increasing or decreasing $r$):
\begin{eqnarray}
ds^2=r^2 \. d\varphi^2 + f(r) \. dr^2 
\end{eqnarray}
Using the squared Lagrangian formalism we immediately get the integrated equation of motion:
\begin{eqnarray}
r^2 \f{d\varphi}{ds}= \te{const}    \label{massa}
\end{eqnarray}
Letting $\theta$ denote the angle between the geodesic in question and a purely azimuthally 
directed line we may rewrite \eq{massa} into:
\begin{eqnarray}
r \. \cos(\theta) = \te{const} \label{fund}
\end{eqnarray}
In particular we see that the tilt of a certain geodesic is completely determined by the radius, 
and that the tilt of the geodesic line increases with increasing radius. 
By considering a thread tightened on the surface we understand that this is very reasonable.

Assuming the rotational surface to be a small perturbation of a cylinder of unit radius $r=1+h$, and assuming the tilt ($\theta$) to be small, we may easily prove from \eq{fund} that to lowest order we have:
\begin{eqnarray}
\f{d^2z}{d\varphi^2}=\f{dh}{dz}
\end{eqnarray}
Thus one may verify that, at least for small velocities and gravitational fields, 
one can explain gravitational attraction by motion on a rotational surface.


\section{Displaying the Earth gravity} \label{earth}
We would like to display why things accelerate at the surface of the Earth. 
We want a clearly curved surface where meters and seconds correspond to 
roughly the same distances as the radius of the cone. We have three parameters 
that determine the shape and size of the embedded surface. Let us therefore 
make three demands, exactly at the surface of the Earth: 
\begin{req}
\sin \Theta_0&=&0.8 \qq &&&\te{The angle of slope for the surface}\\[-1mm]
r_0&=&1 \qq  &&&\te{The embedding radius} \label{re1}   \\ [-1mm]   
\D \t\i{real}&=&1\te{s} \qq &&&\te{The proper time per circumference} \label{te1}
\end{req}
\begin{figure}[t!]
  \begin{center}
      	\epsfig{figure=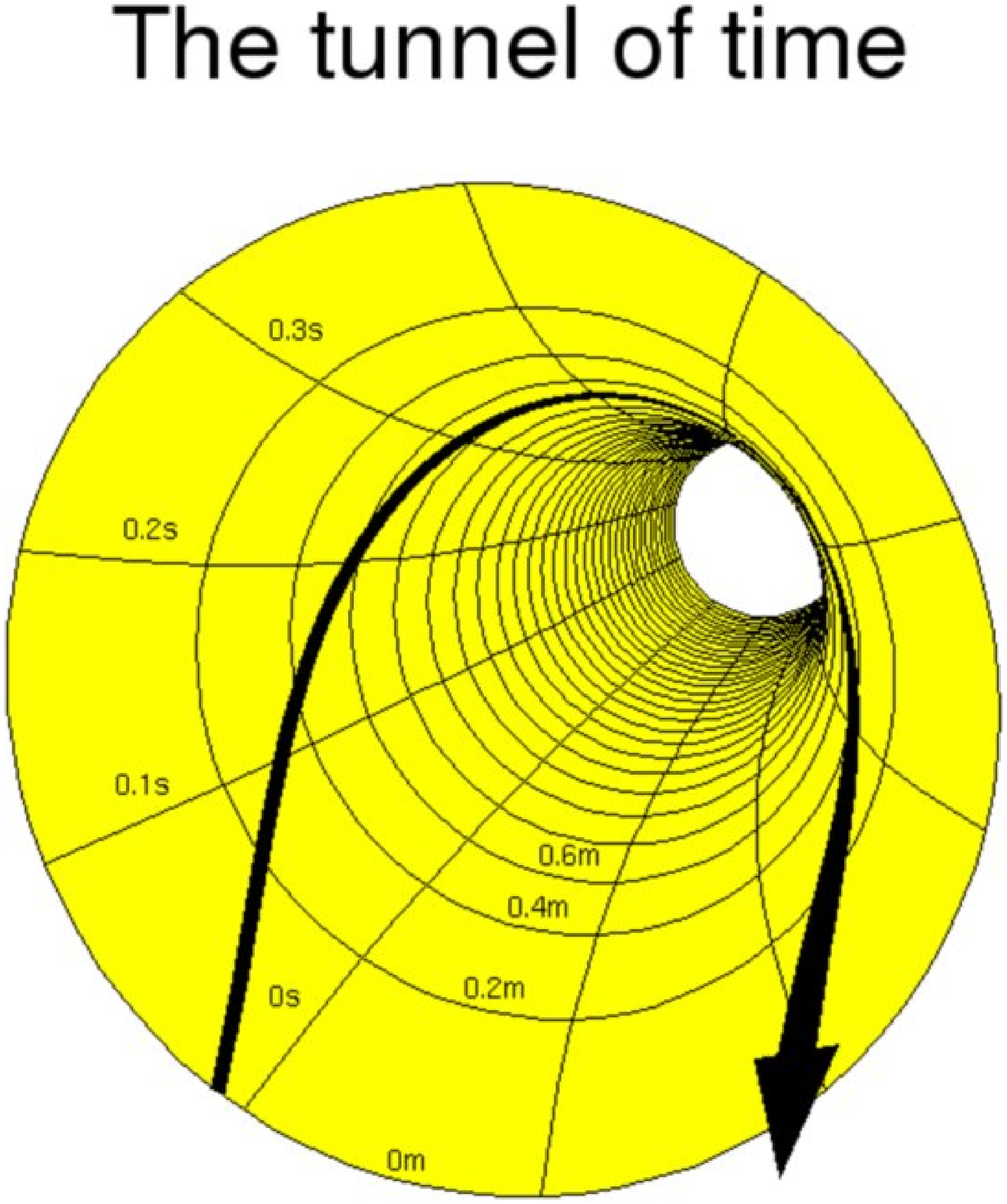,width=7.3cm}
      	\caption{An embedding diagram of the spacetime at the surface 
of the Earth. We see a freefaller moving up and then down again in 
perfect agreement with the Newtonian predictions.}  
     	\label{tunnel}
  \end{center} 
\end{figure}

\noindent
From these requirements it is an easy exercise to find the corresponding values of $k$, $\a$ and $\b$. 
The results are: 
\begin{req}
k&=&\displaystyle \f{\D \t\i{real} c\i{l}}{2 \pi \sqrt{a\i{e0}} R_G} &&&\sim 5.38 \. 10^9 \\
\b&=&\displaystyle \f{a\i{e0}}{1+\(\displaystyle \f{k}{2\te{sin}\Theta_0 \. x_0^2}\)^2} &&& \sim \displaystyle\f{a\i{e0}}{1+\underbrace{4.25 \. 10^{-17}}_\d} \label{beta} \\
\a&=& r_0^2 \displaystyle \f{1}{4x_0^4 \te{sin}^2 \Theta_0+k^2} &&&\sim 1.47 \. 10^{-36}
\end{req}
Here $a_{e0}=a_0(x_0)$ and $c\i{l}$ is the velocity of light. 
Since Matlab only operates at 16 decimals, we cannot cope numerically with the expression for $\b$, 
\eq{beta}. We have to Taylor-expand our expressions for the embedding coordinates $z$ and $r$. 
Limiting ourselves to the exterior metric, and introducing $\D x=x-x_0$ the results are, 
assuming  $0<\D x<<x_0$:
\begin{eqnarray}
r	&\simeq&\f{k \sq{\a}}{\sqrt{\f{\D x}{x_0^2}+\d}} \\
\D z    &\simeq&\sq{\a} \int dx  \f{1}{\f{\D x}{x_0^2} +\d} \sq{1  -   \f{k^2}{4 x_0^4} \. \f{1}{\f{\D x}{x_0^2} +\d}}
\end{eqnarray}
It is now an easy task to show the embedding, see \fig{tunnel}.
I think this picture is really beautiful from a pedagogical point of view. 
Especially pedagogical would it be to make a {\it real surface} in metal, 
using the outside of the {\it trumpet}, with 
meter and second lines drawn on it. Then you can use tightened threads to find 
the time it takes a particle initially at rest to fall say 10 meters. Of course there 
is no {\it action} in this, which kids like, but then again it serves to illustrate that 
there is no action in a spacetime movie, it's a documentary!


\section{Embedding the inside of a black hole}
\begin{figure}[b!]
  \begin{center}
      	\epsfig{figure=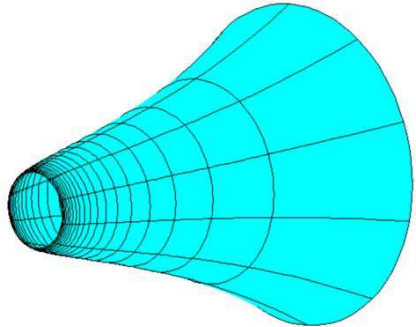,width=5.6cm}
      	\caption{An embedding of the spacetime inside a black hole. 
                 The singularity is to the left and the horizon is to the right. 
                 Schwarzschild time is still the azimuthal angle.}  
     	\label{inside}
  \end{center} 
\end{figure}

Recall the general formulae for the dual metric of a time independent two-dimensional diagonal metric:
\begin{eqnarray}
a=\a \f{a_0}{a_0 - \b} \qq
c=\a \f{-c_0 \b}{\(a_0 -\b\)^2} 
\end{eqnarray}
Inside a the horizon we have negative $a_0$ and positive $c_0$. 
Necessary and sufficient conditions for positive $a$ and $c$ are then:
\begin{reqx}
  0&<&\a \\
a_0&<&\b<0
\end{reqx}
We see that the dual metric exists from the singularity outwards to some $x$ limited 
by our choice of $\b$. Numerically we find that while the metric exists all the 
way into the singularity, the embedding is restricted. An embedding is displayed 
in \fig{inside}. Notice that moving along the trumpet is now timelike motion. In 
particular we see that $t=\te{const}$ is a timelike geodesic 

The trumpet is now narrowing towards the singularity, implying 
acceleration {\it away} from the singularity in the sense that a geodesic that was at some 
point, moving solely in Schwarzschild time, will end up at the horizon. Inside 
the horizon however, moving only in $t$, means moving faster than light. 
Thus it is {\it tachyons} (time-lines) that accelerate out towards the horizon. 

Real observers are however always approaching the singularity. While we 
might find it unintuitive that the shape of the trumpet implies that they 
{\it decelerate} as they move inwards one must consider also the spacing between lines 
of constant $x$. If the distance between $x$-lines is decreasing as we approach the 
horizon, we see that it is quite possible to have increasing $dx/dt$ the closer to 
the singularity that we are, in spite of the trumpet opening up in the ``wrong'' 
direction.


\section{Comments on the two-dimensional analysis}
The dual metric assigns distances between nearby points on the manifold. 
This is what metrics in general do. Under a coordinate transformation the dual 
metric will thus transform as a tensor. 

If we make a coordinate transformation to some wobbling coordinates, the 
dual metric will appear complicated. However, when we {\it embed} it we get the 
static-looking spacetime depicted in \fig{num2}. It's like the ugly duckling 
becoming a swan. The difference is that the coordinate lines will now wind and twist 
on the surface. Notice however that while the embedding of the dual metric is 
coordinate independent there may in principle exist more ways of embedding it 
than as a rotational surface. 

We understand that if the dual metric exists in one coordinate system it 
exists in all coordinate systems. In particular this means that it is sufficient that 
there {\it exists} coordinates where $a$ and $c$ are time independent, for the dual metric to 
exist. Also we realize that, starting from the Schwarzschild line element, written 
in {\it standard} coordinates, we are not excluding any possible dual metrics through 
our choice of coordinates. However we {\it may} be excluding dual metrics by our 
assumption that the dual metric is time independent and diagonal. 

Another small comment. One may think that the dual metric and the original 
metric would have the same affine connection, since the affine connection is all 
that enters the geodesic equation: 
\begin{eqnarray}
\f{d^2x^\l}{d\t^2}+\G^\l_{\m\n}\f{dx^\m}{d\t}\f{dx^\n}{d\t}=0
\end{eqnarray}
Explicitly calculating the dual and original affine connections we find however 
that they are not the same. The explanation is that in the geodesic equation there 
are derivatives with respect to proper distance (eigentime). When we change into 
the dual metric we change the meaning, in a nontrivial way, of the proper 
distance. Then we see the possibility that different affine connections%
\footnote{There {\it should} be some other geometrical object that determines 
geodesics on the manifold that {\it is} the same in both original and dual metric. 
I do not know of it however.}
can produce the same $x(t)$. 

A final comment is in order. The dual metric of \eq{a} and \eq{c} is 
Riemannian from infinity and in to the point where $\b=a_0$. It is however 
geodesically equivalent to the original Schwarzschild metric also {\it inside} this boundary. 
Here $a$ is negative while $c$ is still positive (until we reach the real horizon where 
they both flip sign) implying a Lorentzian signature. This signature boundary 
has nothing to do with {\it coordinates}. Coordinate choices will not affect whether 
there exists negative distances or not. While we can not embed the Lorentzian 
part, it is however interesting to see that we can have smooth geodesics moving 
over something as dramatic as a signature change.


\section{Extension to 2+1-dimensional spacetimes}
I have extended the dual metric analysis to higher dimensions, using 
techniques very similar to those used in the 1+1-dimensional case. Assuming both 
the original and the dual metric to be time independent and diagonal, I find that 
there is in general no dual positive definite metric. The problem is that one needs 
the extra metrical function (connected to azimuthal distance) to fix $x(t)$ for all 
values of the angular momentum $J_0$, but one also needs the extra metrical 
component to get the azimuthal motion right. It's simply over-determined. 

Only for very specific cases of original metrics can we find a dual metric 
that is positive definite. In particular assuming the original metric to have
$g_{tt}=\te{const}$, we find the dual metric simply by flipping the sign of the spatial part. 
This may be understood without any analysis at all. 

However, if we restrict ourselves to demand that only those geodesics that 
correspond to a certain energy ($K$), shall be geodesics in the new metric -- we 
{\it can} find a nontrivial dual metric. In particular studying photons, that have infinite 
$K$, the equations for the dual metric are somewhat simplified: 
\begin{eqnarray}
a&=&\f{\l-1}{k_3 \f{d_0}{a_0}-k_2} \label{t1}\\
c&=&\f{c_0}{a_0} \. a(1-k_2 a) \label{t2}\\
d&=&\l \. \f{d_0}{a_0} \. a \label{t3}
\end{eqnarray}
Here, $k_3$, $k_2$ and $\l$ are gauge constants much like $\a$ and $\b$ in the two-dimensional analysis. 
Consider an original metric of the form:
\begin{eqnarray}
d\t^2=a_0 \. dt^2 + c_0 \. dx^2 -x^2 \. d\theta^2
\end{eqnarray}
Let us assume that $a_0$ and $c_0$ reduces to +1 and -1 respectively at infinity. 
Demanding the dual metric to be flat Euclidean space expressed in 
polar coordinates at infinity, i.e. $a=1$, $c=1$ and  $d=x^2$, 
yields $k_3=0$, $k_2=2$ and $\l=-1$. Inserting these constants into \eq{t1}-\eq{t3} gives:
\begin{eqnarray}
a=1 \label{opt1} \qq
c=\f{-c_0}{a_0} \qq
d=\f{x^2}{a_0}
\end{eqnarray}
This is thus a metric with positive signature that is geodesically dual to 
the original metric with respect to photons.
The spatial part we recognize as the {\it optical geometry}, 
see \cite{optical}. For a brief review applicable to this text see Appendix \ref{optapp}. 

\begin{figure}[b!]
  \begin{center}
      	\epsfig{figure=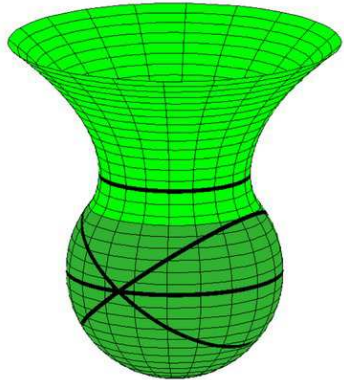,width=4.7cm}
      	\caption{The optical geometry of a plane through a star with $x_0 = 1.2$. 
                 The great circles are closed photon orbits within the star.}  
     	\label{opt}
  \end{center} 
\end{figure}

When $a$ is constant it is easy to understand that geodesics in spacetime are also
geodesics in {\it space}. We may thus embed the spatial part of the dual metric to 
visualize a space where photons move on the shortest path. See \fig{opt}.

It is a little bit fascinating that asking for a metric that is geodesically dual with respect to photons, 
with positive signature, yields the optical geometry, which is normally derived in a 
completely different manner. 

Notice however that we knew in advance that the optical geometry (plus 
time) would be among the possible dual spacetimes. We knew that there existed 
a three-metric (essentially \fig{opt}) for Schwarzschild in which photons moved 
on geodesics. To this space we knew that we could just add time to create a 
Riemannian spacetime in which photons would move on geodesics. Thus we knew 
that the optical geometry (plus time) would be among the possible solutions. It 
turned out to be the {\it only} dual spacetime that reduced to a Euclidean spacetime 
at infinity. 

One may also study the dual geometry that springs from other choices of 
$K$, but then $a$ doesn't become constant and the geometrical information doesn't 
lie entirely in a spatial metric.

Still it is interesting to see that we have a scheme that produces the optical 
geometry for the particular case of photons. What would be really nice would 
be if we, using insights and techniques developed in this paper, could generalize 
the optical geometry to something that makes sense even when the spacetime is 
not conformally static.


\subsection{Comments on the higher dimension analysis}
If we introduce freedoms, like off-diagonality and time dependence, it may 
after all be possible to find a dual metric in higher dimensions. For me it is 
therefore still an open question whether, as soon as we have a fairly nontrivial 
metric, we can decide the exact form of the metric up to a global rescaling 
constant by just studying geodesics? In the two-dimensional analysis it was not 
so, but in the three-dimensional analysis it was so, assuming time independence 
and diagonality of the dual metric. In general however I do not know yet.


\section{Comparison to other works}\label{work}
In Lewis Carrol Epstein's book 'Relativity visualized' \cite{epstein}, there is a similar 
way of visualizing the acceleration towards a gravitational source. The pictures 
illustrated are qualitatively very similar to my own, a star is a bulge on a cylinder 
for instance. In Epstein's view however the azimuthal angle on the rotational 
surface is the {\it eigentime} experienced by an observer (for instance a freefaller). 
The Euclidean length of a curve on the surface is the {\it Schwarzschild} time elapsed. 
All freefallers move on geodesics on the surface. In particular photons, which do 
not experience eigentime, but still travels in Schwarzschild time, are represented 
by straight lines directed along the rotational surface, without the slightest spiral 
in the azimuthal direction. 

The Epstein view is very beautiful in many respects. For instance it 
naturally explains why it takes infinite Schwarzschild time to reach the horizon but 
only finite eigentime. What Epstein is embedding is however not strictly a 
spacetime. A point on his surface is not corresponding to a unique event (in general). 
To see this consider a photon moving in towards the gravitational source and 
then bouncing back outwards. In the Epstein diagram the photon returns to the 
same point that it came from. Thus one point in the Epstein diagram represents 
two events (at least!) in the physical world. Also in the Epstein view one can 
not display spacelike distances. 

Perhaps the biggest advantage of my view, compared to Epstein's view, is 
the opportunity to graphically display how gravity on {\it Earth} can be explained by 
spacetime geometry. This is obviously difficult to accomplish with the Epstein 
view since eigentime and Schwarzschild time are virtually the same thing for us 
Earthlings.

We understand that the two approaches complement each other. It is really 
fascinating however that two such fundamentally different approaches can 
produce more or less the same plots! 

Another way of illustrating curved spacetime is to embed the spacetime in 
a 2+1 Minkowski space. There one has access to null and negative distances. 
See the paper \cite{marolf}, by Donald Marolf. 
The beauty of this scheme is that one can deduce Lorentzian {\it distances} 
between points just from the slope of the surface. Also particles move on 
geodesics, in the sense of shortest Euclidean distance on the surface.%
\footnote{The reason that this solution was not included in my analysis is that 
I only considered time independent dual metrics.}
These  surfaces are not rotational surfaces (in general), and depend on the timelike 
parameter. 

In particular Marolf studies the embedding of a Kruskal spacetime of an 
eternal black hole. The horizons are included in the embedding but not the 
singularities and the infinities. 
A lot of physics can be displayed in this type of embedding. For instance 
one can illustrate how tidal forces become infinite as one approaches the 
singularity. Fascinating.


\pagebreak
\section{Summary}
Always for a two-dimensional time independent diagonal original metric with 
$a_0$ positive and $c_0$ negative we can find a {\it dual} metric where both 
$a$ and $c$ are positive. The new dual metric is dual in the sense that it produces 
the same geodesics as the original metric.
\begin{eqnarray}
a&=&\a \. \f{a_0}{a_0-\b}  \qq \q       0<\b<\te{min}\{a_0\} \qq 0<\a   \label{suma}\\[2mm]
c&=&\a \. \f{-c_0 \b}{\(a_0-\b\)^2} \label{sumc}
\end{eqnarray}
Here $\a$ and $\b$ are {\it gauge} freedoms in the dual metric. 
$\a$ is an overall rescaling. $\b$ is connected to stretching in the $x$-direction, 
while it can also be considered as a rescaling of the original metric. 
In particular, starting from the exterior Schwarzschild 
metric, and demanding that 
the dual metric reduces to Pythagoras at infinity yields $\a=\f{1}{2}$ and $\b=\f{1}{2}$, 
the {\it standard} gauge. The dual line element may then be written:
\begin{eqnarray}
d s^2&=&\f{x-1}{x-2} \. dt^2 + \f{x^3}{(x-1)(x-2)^2} \. dx^2
\end{eqnarray}
In this particular gauge the dual metric stays Riemannian from infinity and in to $x=2$. 
By choosing other gauge constants we can move this boundary arbitrarily close to the horizon. 

We may embed the dual metric as a rotational surface in Euclidean space. 
We have then an embedding freedom $k$. Increasing $k$ means increasing the radius 
and the slope of the rotational surface everywhere. 

\begin{figure}[b]
  \begin{center}
    \psfrag{Observer oscillating around}{Observer oscillating around}
    \psfrag{the center of the star}{the center of the star}
    \psfrag{Observer at rest at}{Observer at rest at}
    \psfrag{the center of the star}{the center of the star}
    \psfrag{Time}{Time}
    \psfrag{Constant time line}{Constant time line}
    \psfrag{Space}{Space}
    \psfrag{mi}{$-\infty$}
    \psfrag{pi}{$+\infty$}
        \epsfig{figure=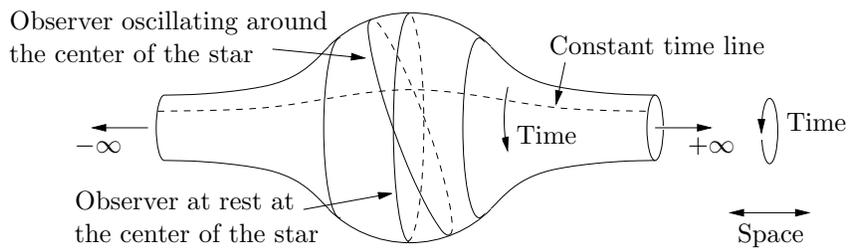,width=10.9cm}
      	\caption{The spacetime of a central line through a star.}  
     	\label{staragain}
  \end{center} 
\end{figure}

A schematic embedding of the dual metric of a radial line through a star is 
depicted in \fig{staragain}. The spacetime surface is layered, so walking around the 
rotational body one lap means that you come to a new spacetime point. Azimuthal 
angle on the surface is proportional to the Schwarzschild time.

\begin{itemize}
\item{
For non-compact stars, using the standard gauge and $k=1$, the difference 
between the radius of the rotational surface and the radius at infinity, is 
proportional to minus the Newtonian potential. 
}

\item{We have all in all three parameters, $\a$, $\b$ and $k$, that affects the shape and 
size of the embedding diagram. In particular this allows us to visualize, 
with substantial curvature, why, and how fast, our keys fall when we drop 
them in our office. 
}

\item{
The dual metric is geodesically equivalent to the original also in regions 
where $a_0<\b$. Here it has Lorentzian signature however. 
}

\item{
We can embed (parts of) the inside of a Schwarzschild black hole simply 
by putting $\b$ negative in \eq{suma} and \eq{sumc}.
}

\item{
In 2+1 dimensions we can not generally find a dual metric that is 
diagonal and time independent. We can however relax the constraints on the 
dual metric to apply, not to all geodesics, but only photon geodesics. That 
way we can re-derive the optical geometry. 
}

\end{itemize}
It would be interesting to generalize the dual metric scheme, to include more 
general original and dual metrics. In particular it would be interesting to study 
an equatorial plane in a Kerr geometry, and restrict ourselves to photons. Also, 
using the dual metric scheme, it remains to be seen if we can somehow include 
the horizon in the embedding, maybe using just a certain set of observers.


\section{Conclusions}
The ideas presented in this article are probably of minor practical use in 
calculations and the finding of new physics. Nevertheless they are, I think, of 
great pedagogical value. They open up our minds to possibilities that we might 
not have considered earlier. This goes for both professionals in the field, but 
even more so for those who have never seen a $g_{\m\n}$, or even an $\eta_{\m\n}$.

For the experts it is probably the concept of the geodesically dual metric 
itself that is most interesting. The signature change in the dual metric may as 
well attract some attention. Also it was nice, though not surprising, to see the 
optical geometry coming naturally from relaxing the geodesic demands to apply 
only to photons. 

For the non-experts it is probably \fig{tunnel}, depicting the spacetime at the 
surface of the Earth, that has the greatest pedagogical value. Especially useful 
would it be to construct such a surface, with meter lines and second lines, and 
threads to pull tight between various spacetime points. Then people get a chance 
to {\it see} how acceleration can be explained by geometry. This I think is very 
powerful, and something that I have longed for, when giving introductory lectures 
on general relativity.


\appendix

\section{Review of the optical geometry} \label{optapp}
Null geodesics are conserved under conformal rescalings of the metric. 
If we have a manifestly time independent metric, with no cross-terms of $dt$ 
(e.g. Schwarzschild), we may rescale it by $g_{tt}^{-1}(x^\m)$ without affecting the null 
geodesics. The rescaled metric will consist of a unit time-time component, and 
a spatial three-metric. It is thus a curved space, with time. There is no 
acceleration or time dilation. For such a metric, a so called ultrastatic metric, it is easy 
to understand that geodesics in spacetime are also geodesics in space. 

So, by rescaling the spatial part of, for instance Schwarzschild, with $g_{tt}^{-1}$ 
we create a {\it space} in which photons moves on geodesics. This rescaled space is 
known as the optical space.


\clearpage


\begin{thebibliography}{999}
\bibitem{marolf}Marolf, D. (1999). {\it Gen. Rel. Grav.} {\bf 31}, 919 

\bibitem{epstein}Epstein, L. C. (1994). {\it Relativity Visualized}, (Insight Press, San Fransisco), 
ch. 10,11,12

\bibitem{weinberg}Weinberg, S. (1972). {\it Gravitation and Cosmology: 
Principles and Applications of the General Theory of Relativity}, (John Wiley \& Sons, U.S.A), p. 77

\bibitem{optical}Kristiansson, S., Sonego, S., and Abramowicz, M. A. (1998). 
{\it Gen. Rel. Grav.} {\bf 30}, 275

\bibitem{dinverno}D'Inverno, R. (1998). {\it Introducing Einstein's Relativity}, 
(Oxford University Press, Oxford), p. 99-101

\end{thebibliography}
\end{document}


\noindent
{\Large \bf Addendum to ``Embedding Spacetime via a Geo\-desically Equivalent 
Metric of Euclidean Signature''} 
\\[5mm]
Rickard Jonsson%
\footnote{Department of Theoretical Physics, Physics and Engineering
Physics, Chalmers University of Technology, and G\"oteborg University, 412
96 Gothenburg, Sweden.\\ E-mail: rico@fy.chalmers.se. Tel +46317723179.}
\\
\\
{\bf Abstract.}
I extend the analysis of Jonsson, R. (2001). {\it Gen. Rel. Grav.} {\bf 33}, 1207,
concerning the shape of an embedding of the dual spacetime of a line through a planet of 
constant proper density. In particular I find that in the non-compact limit, 
the embedding of the  interior of the planet can be chosen to be spherical.

\section{A spherical interior dual metric} \label{csphere} 
We know that in Newtonian theory, a particle that is in free fall
around the center of a spherical object of constant density, will
oscillate with a frequency that is independent of the amplitude. This
would fit well with an internal dual spacetime that is {\it spherical}. We
saw in \cite{dual} the possibility to choose parameters $\beta$ and $k$
that produces substantial curvature of the embedding also for the case of the 
weakly curved spacetime outside our Earth. It is therefore natural to ask whether we can
find parameters $\beta$ and $k$ such that the internal geometry becomes spherical. In
the following we will see that this is not generally the
case but one can choose parameters such that it is exactly spherical
in the non-compact limit.

\subsection{Conditions for spheres}
\begin{figure}[ht]
  \begin{center}
    \psfrag{r}{$r$}
	\psfrag{z}{$z$}	
	\psfrag{R}{$R$}
      \epsfig{figure=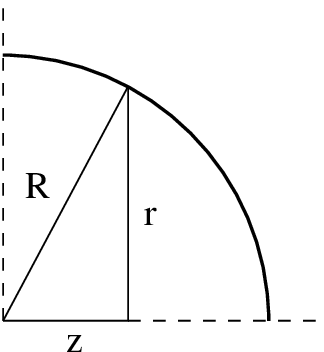,width=3cm,angle=0}
      	\caption{Definitions of variables for a sphere. The axis of
  rotation (around which time is directed) is the horizontal axis.}  
     	\label{f1}
  \end{center} 
\end{figure}

For a sphere of radius $R$, we introduce definitions of $r$ and
$z$ according to \fig{f1}.
The Pythagorean theorem gives $R^2=z^2+r^2$. Differentiating this relation
with respect to $z$ and again using 
the Pythagorean theorem yields:
\begin{eqnarray}\label{sphere}
	\f{dr}{dz}=-\sq{\(\f{R}{r}\)^2-1}
\end{eqnarray}
For a general surface of revolution described by functions $z(x)$ and $r(x)$, with a corresponding 
metric on the form $ds^2=a(x) dt^2+c(x) dx^2$, we have according to Eq. (23) of \cite{dual}:
\begin{eqnarray}\label{genrot}
	\f{dz}{dx}=\sq{c(x)-{r'}^2}
\end{eqnarray}
Here a prime indicates differentiation with respect to $x$, as it will henceforth. 
From \eq{sphere} and \eq{genrot} we readily find:
\begin{eqnarray} \label{lr}
	R^2=\f{r^2}{1-{r'}^2/c}
\end{eqnarray}
This must hold, for a fixed $R$, for all $x$ within a certain interval in order for 
the rotational surface to be spherical in that interval.

Note that the term ${r'}^2/c$ equals $dr^2/ds^2$ where $ds$ is 
the infinitesimal distance measured along the rotational curve $r(z)$. 
Thus ${r'}^2/c$ is a measure of the slope of the rotational curve.

It is a short exercise to integrate the slope given by \eq{lr} in terms of $R$ and $r$ 
(also using $ds^2=dr^2+dz^2$) to find that, for a fixed $R$, \eq{lr} implies a circular 
rotational curve (with center on the $z$-axis). Thus \eq{lr}, for a fixed $R$, is a both 
sufficient and necessary condition for the rotational surface to be spherical.

\subsection{The dual interior metric}
Taking the square of Eq. 21 in \cite{dual} gives:
\begin{eqnarray}\label{ex1}
r^2=\a k^2 \f{a_0}{a_0-\beta}
\end{eqnarray}
Taking the derivative of this expression with respect to $x$ and using the expression 
for $c(x)$ given by 
Eq. (11) of \cite{dual} as well as the explicit expressions for the Schwarzschild interior 
metric (inside the planet) given by Eqs. (30) and (31) of \cite{dual} gives: 
\begin{eqnarray}\label{ex2}
	{r'}^2/c= \f{\beta k^2 x^2}{4 x_0^6 (a_0-\beta)}
\end{eqnarray}
Inserting Eqs. \eqref{ex1} and \eqref{ex2} into \eq{lr} we readily find that for the embedding 
of the internal dual metric to correspond to a sphere of radius $R$ we must have, 
for all $x$ where $-x_0<x<x_0$:
\begin{eqnarray}\label{hattr}
R^2 (a_0-\beta)- R^2   \f{\beta k^2 x^2}{4 x_0^6} -\a k^2 a_0=0 \label{mast}
\end{eqnarray}
For \eq{hattr} to be true for all $x$ in the interval in question, it must be true for every
power in $x$. We note that $a_0$ has a lot of powers in $x$ which
would mean that the sum of the factors multiplying the $a_0$-terms would have to
vanish. But then the $x^2$-term in the middle cannot be canceled by anything. 
Thus the expression {\it cannot} be true for all values of $x$ in the interval in question,
and thus the interior dual metric is not exactly spherical. 

\subsection{Approximative internal sphere}
We {\it can} however produce something that is very similar to a sphere. First, to be specific, 
we let $\a=(1-\beta)/k^2$ meaning unit radius at infinity. 
We then expand the two occurrences of $a_0$ in \eq{mast} to second order in $x^2$. 
Demanding the equation to hold to zeroth, first%
\footnote{It is always satisfied to first order.}
 and second order in $x$ yields after some simplification (in Mathematica):
\begin{eqnarray}
\beta&=&\f{a_{00}\.(R^2-1)}{R^2-a_{00}}  \q \te{where} \q a_{00}\equiv a_0(0)=
\f{1}{4}\(3\sq{1-\f{1}{x_0}}-1\)^2 \label{h1} \\  
k&=&\f{2 x_0^{7/4}}{\sq{3\sq{x_0-1}-\sq{x_0}}}\label{h2}
\end{eqnarray}
Notice that \eq{h2}, coming directly from the second order demand, is independent of $\beta$. 
The first relation is nothing but Eq. (21) in \cite{dual}, where one has demanded 
$r=R$ and $\a=(1-\beta)/k^2$, taken at $x=0$.

Inserting $\beta$ and $k$ from Eqs. \eqref{h1} and \eqref{h2} into a program that plots an embedding of 
the dual spacetime yields pictures with 
a very spherical appearance. The way it works is that the center of the bulge has the exact 
radius and curvature of a sphere, then the rest is not exactly spherical.

\subsection{Spheres in the Newtonian limit}
\begin{figure}[ht]
  \begin{center}
      \epsfig{figure=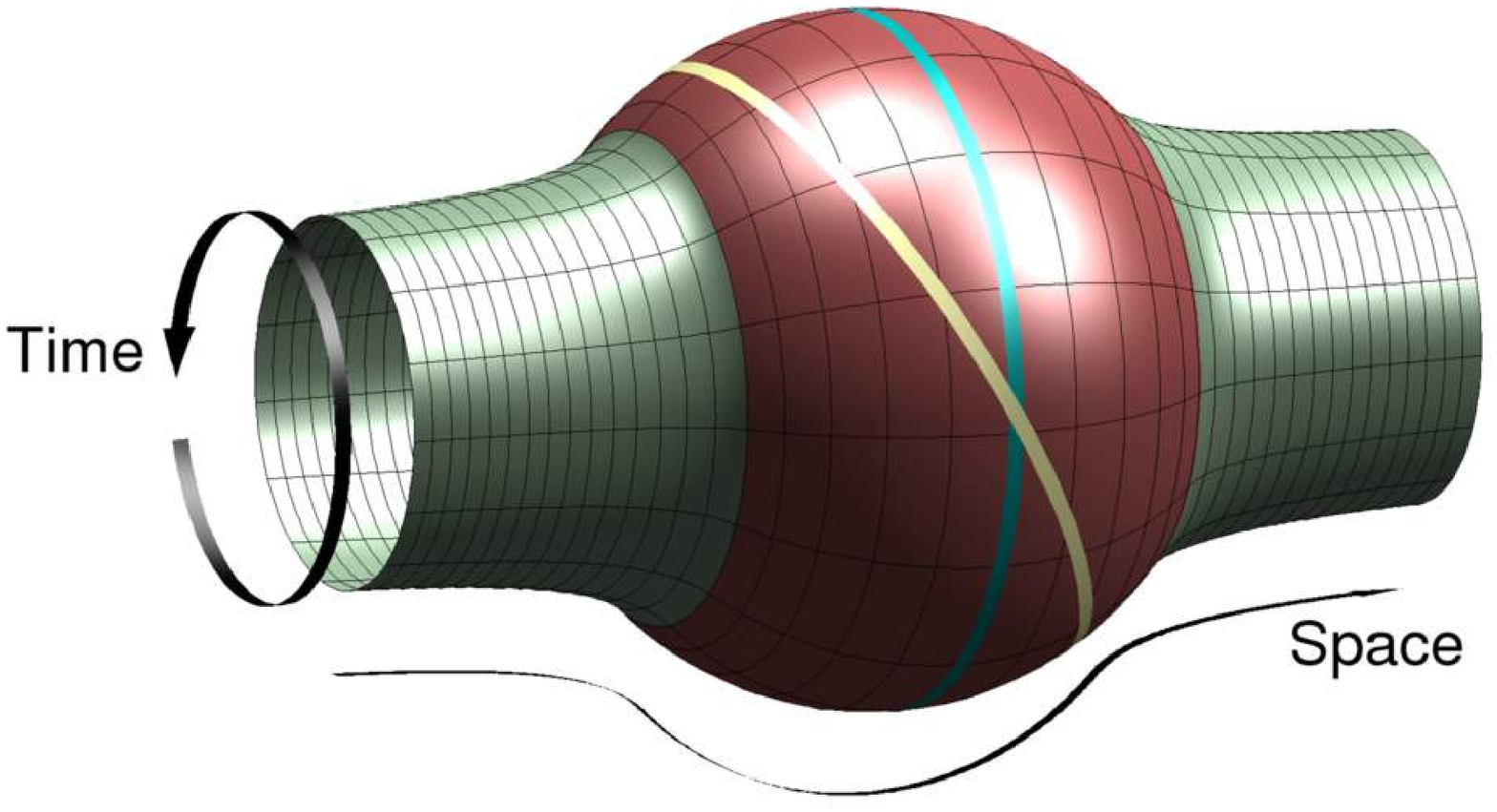,width=10cm,angle=0}
      \caption{An embedding of the dual spacetime of a central line through the Earth.
Note the worldlines of the two freefallers on the interior sphere. One is at rest in the center, 
the other is oscillating about the Earth's center. The time direction is chosen opposite to that illustrated in Fig. 8 of \cite{dual}, to better connect to ordinary flat spacetime diagrams.}  
     	\label{f2}
  \end{center} 
\end{figure}

From \eq{lr} we have for a sphere:
\begin{eqnarray}
\f{{r'}^2}{c}=1-\f{r^2}{R^2} \label{lr3}
\end{eqnarray}
Using the specific expressions for $\beta$ and $k$ given by Eqs. \eqref{h1} and \eqref{h2}, 
together with Eqs. \eqref{ex1} and \eqref{ex2}, we may evaluate both the left and the 
right hand side of \eq{lr3}.
The left hand side is then a measure of the actual slope of the embedded surface 
(for the $x$ in question), whereas the right hand side is a measure of the slope 
the embedding {\it should} have (for the $x$ in question) in order to correspond to a sphere.
Expanding the expressions on both sides of \eq{lr3} to lowest non-zero order%
\footnote{Terms like $x^2/x_0^3$ are of course treated as $1/x_0$-terms.}
 in $1/x_0$, we find that the equality holds {\it exactly} for all $x$.
Thus in the limit as $x_0 \rightarrow \infty$ we get an exact sphere. 
For the case of the Earth (approximating it to be of constant proper density) 
we have $x_0=7.19\.10^8$. Thus one might guess that the higher order ($1/x_0^2$)
differences between the actual slope and the slope of a perfect sphere, are rather small.
An embedding of the Earth spacetime is displayed in \fig{f2}. Here I have chosen the 
radius of the central bulge to be twice the embedding radius at infinity, thus choosing $R=2$. 
The parameters $\beta$ and $k$ are given by Eqs. \eqref{h1} and \eqref{h2}. The rotational radius $r(x)$ 
is given by Eq. (21) of \cite{dual} 
and $z(x)$ is given by numerical integration of Eq. (25) of \cite{dual}.
As can be seen in \fig{f2} the bulge is quite spherical.

\subsection{The spacetime circumference}
The dual distance around the spacetime rotational surface is 
given by $\Delta s=2\pi r=2\pi k \sqrt{a}$. 
Letting $\Delta t$ denote the dimensionless and rescaled coordinate time per circumference, 
we have also $\Delta s=\sqrt{a} \Delta t$. It follows that: 
\begin{equation}
\Delta t=2 \pi k
\end{equation}
Furthermore, the real Schwarzschild coordinate time $\Delta t\i{real}$ 
is related to the dimensionless rescaled Schwarzschild time via 
$\Delta t=\Delta t\i{real}/(2MG/c\i{l}^3)$. 
Here $c\i{l}$ is the velocity of light. It follows that around the embedded dual 
spacetime we have:
\begin{equation}\label{del}
\Delta t\i{real}=4 \pi k \frac{ M G}{c\i{l}^3}
\end{equation}
For large $x_0$, \eq{h2} can be approximated by $k\simeq \sqrt{2} x_0^{3/2}$. Also, we have 
$x_0=R\i{E}/({2MG/c\i{l}^2})$ where $R\i{E}$ is the radius of the Earth. 
Inserting this into \eq{del} gives:
\begin{equation}\label{del2}
\Delta t\i{real}\simeq 2 \pi \sqrt{\frac{R\i{E}^3}{MG}}
\end{equation}
Inserting the appropriate constants yields a circumference of roughly 84 minutes.

For the case of a spherical bulge -- the period of revolution corresponding to a geodesic 
circling the spacetime bulge, equals the circumference of the spacetime.
It is a short exercise of Newtonian mechanics to show that the period of revolution for an
object oscillating about the center of the Earth (within the Earth) is given by 
(assuming the Earth to be of constant density):
\begin{equation}
\Delta t\i{Newton}= 2 \pi \sqrt{\frac{R\i{E}^3}{MG}}
\end{equation}
Comparing this with \eq{del2} which holds to good accuracy for 
non-compact planets such as the Earth, we have a perfect match.